\documentstyle[preprint,aps]{revtex}
\tightenlines
\include{psfig}

\begin{document}
\draft
\preprint{ }
\title{Properties of the energy landscape of network models for 
covalent glasses
}
\author{J. Christian Sch{\"{o}}n} 
\address{Institut f{\"{u}}r Anorganische Chemie,
Universit{\"{a}}t Bonn\\
 Gerhard-Domagk-Str. 1, D-53121 Bonn\\
  Germany 
}
\author{Paolo Sibani} 
\address{Fysisk Institut,
Odense Universitet\\
 Campusvej 55, DK-5230 Odense M\\
 Denmark
} 
\date{\today}
\maketitle
\begin{abstract}
We investigate the energy landscape of two dimensional network models for 
covalent 
glasses by means of  the lid algorithm. For three different particle densities and 
for a range of network sizes, we exhaustively analyse many configuration space 
regions  
enclosing   deep-lying energy minima. We extract the local densities of states and 
 of minima, and the number of states and minima accessible below a 
certain energy barrier, the 'lid'. These quantities show on   average a close
 to exponential growth as a 
function of their respective arguments. We calculate the configurational entropy 
for these pockets of states and find that the excess specific heat exhibits a peak at 
a critical temperature associated with the exponential growth in the local density 
of states, a feature of the specific heat also observed in real glasses at the glass 
transition.
\end{abstract}
 
\pacs{61.43.-j,61.43.Fs,5.40.+j,5.70.Fh}


\section{Introduction} 
Since the beginning of the last decade,  systems with complex  multi-minima 
energy landscapes have attracted increasing 
attention\cite{Landscape_paradigms}, with a  common theme
being thermal relaxation or more generally, stochastic dynamics on the landscapes. 
Such  dynamics can  either have  intrinsic physical interest   or  
 be utilized as an optimization device as done  in annealing techniques.   
A number of approaches have been developed, focusing on 
different aspects of the problem: On the one hand, molecular dynamics and Monte 
Carlo simulations are   performed, often using highly refined model potentials, 
which are designed to reproduce as closely as possible the actual dynamics
 at short times \cite{Olig95}
and the equilibrium statistical mechanical properties of the system 
\cite{Olig97,Ciccotti87,Vashishta90}, respectively. On the other hand, one uses 
simple models describing only selected features of the system, which are amenable to analytical 
techniques \cite{Hoffmann88,Bouchaud97}  or can be studied numerically 
\cite{Kremer88} in enough detail to yield general insights into the qualitative and 
semi-quantitative behaviour of the system. As part of the latter approach one can 
consider abstract graph models, which formally can be thought of as "lumped" 
representations of the energy landscape itself
\cite{Hoffmann88,Sibani93,Schon96}. 

The network models for covalent glasses presented in this paper belong to the 
second class of approaches, since they are tailored to  describe the slow part of the complex
hierarchy of relaxational degrees of freedom\cite{Schon97,Elliott90,Gutzow95}
which characterize glasses: On 
the shortest  time scales, we are dealing with small vibrations in the immediate vicinity of 
individual minima of the energy surface. These are responsible for most of the 
vibrational and reversible elastic properties of the glass. Here, the analysis usually 
employs matrix diagonalisations at the point of the minimum, or short time MD-
simulations. At the next level, neighbouring minima are accessible 
by crossing very small barriers. This mechanism is probably responsible for some 
of the anomalous low-temperature properties of glasses. One would suspect that, 
at this level of detail,  the so-called "two-state-models" 
\cite{Phillips72,Anderson72} and their 
descendants e.g. the soft potential models \cite{Galperin85,Buchenau91}, would be 
a relevant theoretical 
description, which can be complemented by studying the diffusion of (single) 
particles by MD/MC-simulations at various temperatures. 
At large time scales and/or high temperatures (up to the point where the glass 
melts), the main structural feature of the glass is its topology.
Accordingly,  one usually 
visualises the glass as a random network of  building units 
\cite{Elliott90,Gutzow95,Zallen83}, 
where the links  represent either chemical bonds (e.g. $Si-Si$) or sequences of 
chemical bonds (e.g. $Si-O-Si$ or $B-O-B$).
The relevant excitations are 
 likely  to be long wavelength  distortions of the covalent network, which involve 
 the displacement of many atoms, with each displacement small   
 compared to  the interparticle distance.
Such  distortions  can substantially change the geometry of the 
structure, while they  only weakly affect its topology. 
In systems containing thousands of atoms per 
simulation cell   it becomes 
computationally  very expensive if not impossible to run MD/MC-simulations for 
the required  times while still  
using highly refined potentials.  However, the structural and energetic hierarchy of a covalent 
glass, leading to a separation into vibrational, geometric, and topologic properties 
of the glass, opens the possibility of employing network models on lattices to 
selectively describe the topology of the glass. Similar lattice models for polymers 
have been successfully analysed in recent years using MC-simulations 
\cite{Kremer88}.

A salient feature of glasses is the glass transition, with a peak of the (excess) 
specific heat capacity $C_p$ at a temperature near the transition temperature 
$T_G$ 
\cite{Elliott90,Gutzow95}. This peak is usually associated with the so-called 
configurational 
entropy reflecting the multitude of different topological structures accessible to the 
glass during this transition. Each configuration represents a basin around a 
relatively stable local minimum of the potential energy. Thus, the configurational 
entropy is an excess entropy of the glassy state relative to the crystalline state, 
whose entropy at this temperature is dominated by the vibrational states. Based 
on the previous discussion, it should therefore be possible to link the excess 
entropy of glasses to statistical features of the energy landscape of  covalent 
network models. 

Motivated by these considerations we numerically analyzed  the energy landscape of small network 
models on two-dimensional lattices for a range of sizes and densities. The 
restriction of the nodes of the networks to a lattice allows complete 
characterization of subsets  of the discrete landscape.
Using the lid algorithm \cite{Sibani93}, we performed  exhaustive searches of local 
regions  around deep-lying minima (so-called pockets), which yield information on the local 
densities of states, the available configuration space volume and the distribution of 
neighbouring minima. This information is then used to understand some of the 
features of the thermodynamics and dynamics of the system.
 
\section{ Model and algorithm} 
\subsection { Lattice network model} 
The networks were placed on square lattices with periodic boundary conditions. 
The size of the repeated cell $S \times S$ ranged from $10 \times 10$ to $20 
\times 20$ grid points. The 
 number density $\rho = N_A/V$, $V = S^2$, of the  $N_A$ building units per cell (
 such units are henceforth for 
simplicity  called atoms)  was chosen to be approximately $0.13$, $0.14$ and $0.15$. 
 The interaction potential given by Eqs.~\ref{radial} and \ref{angular})    between the atoms consisted of a sum of a 
two body and a three body term. The former, $V_2(r)$, grows quickly towards large positive 
values for distances $r < 1.8a $ and equals infinity for $r < a$, while for $r > 3.2a$ it smoothly approaches 
zero. The lattice parameter 
$a$ was chosen in such a way that the optimal distance between atoms was 
about $2.1a$.   The three body term  $V_3(\theta)$ details the angular dependence of
the interactions among nearest neighbors
(it only applies for  $r < 3.2a$). It has a 
minimum at about $120^o$,  reaches infinity for angles smaller than $80^o$ , and 
vanishes smoothly when the angle approaches  $180^o$. The actual formulae used are not 
very important,
but are nevertheless mentioned for completeness:  
\begin{equation}
V_2(r) = \left\{ \begin{array}{r@{\quad:\quad}l} +\infty & r < 1.6 \\
 1/3(r-1.6)(r-3.2)^2 (r-7.1) & 1.6 \le r \le 3.2 \\ 0 & r > 3.2 \end{array} (1a) \right.
 \label{radial}
\end{equation}
and
\begin{equation}
V_3(\theta) = \left\{ \begin{array}{r@{\quad:\quad}l} +\infty & \theta < 80^o \\ 
5.8 \times 10^{-8} (\theta - 80^o)(\theta - 180^o)^2 \theta & 80^o \le \theta \le 
180^o \\ 0 & \theta > 180^o \end{array} (1b) \right.
\label{angular}
\end{equation}

Thus, four-fold coordination was possible, but not favoured energetically. A typical 
metastable configuration is shown in fig.~1. The binding energy of such deep-lying 
minima was in the range of $-6.0$ to $- 6.5$ eV/atom. The (crystalline) ground 
state of an infinite system - without an underlying lattice - has a hexagonal structure 
similar to that of a graphite layer, with each atom surrounded by three equidistant 
neighbours. In the finite on-lattice system, most of the deep-lying minima 
encountered contained a certain amount of adjacent slightly flattened hexagons, 
irrespective of the density, with the flattening due to the underlying square lattice. 
Note that,  owing to the periodic boundary conditions, there will be configurations
which appear to be different, but are equivalent to one  another as  they are 
connected by translations and/or rotations of the system. In the present work,
such configurations are identified as equivalent and   counted only once.

Another issue that needs to be considered when placing the networks on a lattice is 
the mesh size dependence of the results,  which should of course be
 negligible. Clearly, halving the lattice constant will lead to more states. 
 But as long as the width of the energy interval $\delta E$ used in the 
investigation
  is such that these new configurations all lie 
within the  interval $[E,E+\delta E]$, the change in the lattice parameter 
just adds a constant to the entropy. This corresponds to a parallel shift in
 a semi-logarithmic plot of the local density of states, 
and, as we shall see,  does not affect our conclusions. 
We have ensured that this requirement is fulfilled reasonably well for our 
 on lattice networks.
Finally, one has to establish  the connectivity of the configuration space.
In our case, the neighbours of a given configuration are obtained by 
all possible moves of  a single atom from its current position to one
of the neighbour points of the lattice. With $N_A$ atoms
we have in a $d$ dimensional square lattice $2 d N_A $ neighbors for each configuration. 
There is of course an amount of arbitrariness in the choice of the elementary moves
 which  define the connectivity of the landscape. We were guided by the simple
 physical consideration that a single move should involve a change of coordinates
 which is  small, i.e. of order $a$. This seems reasonable considering that  
 collective moves are associated to vibrational motion, and thus take place
 on time scale much shorter than those of interest here. 
 
\subsection{ Lid algorithm} 
The method chosen to investigate the energy landscape of the networks is the  
so-called lid algorithm. Here, we will only give a short summary of the basic procedure; 
a more detailed description of the algorithm and the problem-independent 
implementation we have used can be found in the literature 
\cite{Sibani93,Sibani98}. 
Note that the lid algorithm is only applicable for discrete configuration spaces; 
continuous energy landscapes require a modified approach, the so-called threshold 
algorithm \cite{Schon96,Schon96b}.

The central idea is to restrict  the investigation of the 
configuration space  to a set of smaller subregions, called pockets, which 
surround local energy minima. 
These pockets contain a few hundred to a few million states and can be explored 
exhaustively. The procedure is as follows: Starting from a minimum $x_i$, we list 
all the states that are accessible to a nearest neighbor 
random walk,  which starts at $x_i$ and which is restricted to
states of energy lower than  a prescribed energy
value  $L_k$, henceforth called the `lid'. (By nearest neighbor random walk we 
mean a random walk in configuration space, 
whose steps consist of  precisely the previously defined  elementary moves). 
From the exhaustive listing, we can compute the
the local densities of states $D(E;L_k,x_i)$ and the local density of 
 minima  $D_M(E;L_k,x_i)$ for states available below the lid $L_k$.
 Integrating $D(E;L_k,x_i)$ with respect to $E$ over the interval $[0,L_k]$
 we obtain the number of available states $N(L_k;x_i)$, and by the same
 operation applied to $D_M(E;L_k,x_i)$ the number of available local minima 
$M(L_k;x_i)$.  The search is repeated for successively higher lid values $L_{k+1}$,
 $L_{k+2}$,  etc., and for as many other pockets as possible. Note that the pocket is delimited 
by energy barriers instead of some distance in configuration space from the 
starting minimum. This is a natural first choice of selecting a physically relevant 
region of configuration space, due to the presence of 
an  Arrhenius factor  
$\tau_{esc} \propto \exp(L_k/T)$, in the   time of escape out of the pocket. 
One would like to repeat the analysis starting from every local minimum 
within a given pocket. Since the number of local minima within a given pocket of 
decent size ($> 10^5$ states) was already on the order of $10^4$,
this was not possible, and the available computer time was instead used for a 
sampling over more pockets for a given cell size and density. The starting minima 
for these pockets were found by using either simulated annealing or an adaptation of the lid 
method itself.
 
\section{Results}
In this section we first describe  in some detail the 
 features of the data obtained by the exhaustive investigations, 
and then discuss   their links to  the 
thermodynamical features of the model and of glasses. 
 
\subsection{General features of the data}
Figures 2a, 2b and 2c  describe three different pockets,  belonging 
to the systems $(N_A = 22; S = 12)$, $(N_A=36;S=16)$ and $(N_A = 53; S=20)$ 
respectively.
In each case we plot, on a semilogarithmic scale, 
$N(L)$ and  $M(L)$ as functions of the lid $L$, and 
$D(E;L_{max})$ and $D_M(E;L_{max})$ as function of the energy $E$.
For convenience, the dependence on $L_{max}$ of the last  two functions  
is  left understood.
One notices the average 
exponential growth in all four quantities, with a flattening of $D(E)$ and $D_M(E)$ 
for 
energies near the maximum lid. Such a behaviour is exhibited in a large majority of 
the pockets, independent of (linear) size $S$ and density $\rho$.

Figure ~3 shows semi-logarithmic plots of the local density of states $D(E)$
 for five different pockets belonging to systems of  
 different cell sizes $S$ and different densities $\rho$ as indicated in the
 caption.
 These data are meant to 
illustrate the  variations in the shape of 
$D(E)$ seen in the simulations. Note that the larger systems shown in 
the panel a)  have a  steeper
growth of the local density of states than the smaller systems displayed in panel b).
 In the following we denote
by   $x_i$ the state of lowest energy   within  
a certain pocket, and similarly   by the index $i$ various other
quantities associated with the   pocket.
In about 60\% of all pockets studied  the curves for $D(E)$ resembled those shown 
in (a1) (26\%) or (a2) (34\%), i.e. they exhibited simple exponential growth 
$D(E) = g_i \exp(E/E_i) = g_i \exp(\alpha_i E)$, 
with $g_i = 1$ and $g_i > 1$, respectively. Here, $E_i = 1/\alpha_i$
 is  the energy scale  characterizing  the exponential growth laws. 
  In addition, the closely related case (b1) appeared for 29\% of the pockets, 
while (a3) and (b2) represent some less common examples that occured in 3\% and 
8\%  of all cases, respectively. In some instances 
((b1) and (a3)), the curves are best described by splitting them into two sections, each 
showing exponential growth, but with different `growth factors' $\alpha_i (1)$ and 
$\alpha_i (2)$.

Figure 4 is a plot of   $\alpha =<E_i>^{-1}$ versus the linear size of
the cell, for different densities.  Here the
averaging is performed  over  all the pockets of systems with the same 
density $\rho$ and cell size  $S$.
 It is clear that $\alpha$ increases and correspondingly
 $<E_i>$ decreases with decreasing density and increasing cell size. This result agrees 
qualitatively and to a certain extent also quantitatively with what one would 
expect from the  simple free volume analysis 
\cite{Schon97}  outlined in the appendix. To show the agreement, 
 the calculated curves 
based on eq.\ref{A7} for $\rho = 0.13$, $0.14$ and $0.15$ 
are also depicted in figure 4
together with the actual data. 
 
For a given value of $N_A$ and $S$ (considering only the larger pockets), 
one finds 
a considerable spread in the values of $E_i$, which vary up to a factor of two. This 
is 
quite different from the corresponding results obtained e.g. for 
spin glasses\cite{Sibani94,Sibani98b}, but it is 
consistent with the fluctuations of $N(L)$ and $D(E;L)$ around the average 
exponential growth curve. The size of the pockets and their local $DOS$ are quite 
variable, possibly because different spatially localized excitation patterns of the 
network can exhibit different growth behaviour in the associated $DOS$. Thus 
large 
side-basins with different growth laws can appear upon exceeding some energy 
barrier.

The density of minima shows a strong similarity to the density of states. Again, 
exponential growth is found, with the ratio of the growth factors 
$\alpha_i(D)/\alpha_i(D_M)$ mostly in the range 1-2. This result agrees with the 
observation that the number of accessible states in a pocket $N(L)$ is more or 
less 
proportional to the number of minima $M(L)$, with the ratio 
$N(L_max)/M(L_max)$ 
mostly in the range $15-60$.

 As a function of $L$, $N(L)/M(L)$ increases only slightly with the lid value. 
This holds true for all pockets investigated; and this observation is reflected in 
$M(L)$ running nearly parallel to $N(L)$ in a semi-logarithmic plot, once a lid 
value is 
reached where the first side-minima are accessible. Of course, $D_M(E;L_{max})$ 
eventually decreases,  
in most instances for energies close to the lid, $E \approx L_{max}$.  

It is natural to investigate to what extent the properties of a pocket 
 depend on the energy of its  lowest energy state. Denoting as the depth of a 
pocket the smallest energy barrier which must be crossed in order to gain access 
to a lower minimum, we find that at least for larger pockets,   $(N(L_{max}) > 
1000)$,  
the depth grows as the energy of the lowest state decreases. With regard to the 
various growth factors, there exists a weak trend, insofar as pockets around low-
energy minima tend to grow on the average slightly faster than the high-lying 
pockets, i.e. they have lower values of the $E_i$'s.
 However, many smaller 
pockets, often containing less than a dozen states and only a couple of minima, are 
interspersed among the larger ones along the energy-axis. Their depths do not 
appear to follow any particular pattern, but their growth quite closely follows an 
exponential law, analogous to the larger pockets.

\subsection{More detailed description of a typical example}

Let us consider in some more detail a typical case 
such as the system with  $(N_A =39, S =17)$.
 There are many  local minima, each 
identified by its depth and by  the energy of its lowest state. 
Each valley has its local density of states, which only includes 
states accessible by crossing energy barriers lower than the
depth. All these local densities of states are to a good approximation
exponential, and thus characterized by  a growth 
rate, and its inverse $E_i$.   Figure  5 shows a set of  $E_i$'s  
(circles) and of   depths (squares)   plotted as function of the
energy $E$ of the lowest state in the corresponding pockets.
  We see that the inverse growth rates do not vary much from 
  pocket to pocket. The   average value of $E_i$ is for this system 
$ E_{gr}^{av} \approx 0.035 \pm 0.005$ eV/atom. On the other hand,
we note  that pockets surrounding states of very low energy tend to be deeper than
those surrounding less deep minima. In other words, the lower the energy
the more rugged the landscape appears to be.

 In addition, we 
show in fig. 6 for the "ground state"- pocket ( "ground state" = deepest minimum 
found on the energy landscape) the differences $\delta DOS(E;L_k)$ between the 
densities of 
states for subsequent lid values $L_{k-1}$ and $L_k$.
 We note that these densities of states belonging 
to the added sub-regions at lid L are very similar up to the point of joining the main 
pocket. In particular, we note that at the lid $L=0.14$ eV/atom
 a large group of similar 
basins that are about as deep as the starting minimum join the main pocket. If 
one considers the growth factors of these $\delta DOS$, one finds 
$E_{i}^{av}(\delta DOS) \approx 0.027 \pm 0.03$ eV/atom. These values are quite 
similar to 
the growth factor of the whole 
"ground state"- pocket $(E_i \approx 0.028)$ eV/atom,
 and lie at the lower end of 
the  range of growth factors shown in figure 5. 
Thus, the larger sub-basins added with increasing lid-size 
are quite similar to the pockets encountered during the sampling of the whole 
energy landscape.

\subsection{\bf Configurational entropy}
As mentioned in the introductory section, it is reasonable to discuss the excess 
entropy of the glass in terms of configuration space properties of network models. 
We have to assume that, at the temperatures of interest, the model system would 
thermalize in the pockets described by our numerical investigation.  In 
this context, it is important to realize that the system experiences a qualitative 
change of behaviour at the temperature $T = E_i$. Analyzing the expectation value 
of the energy of a pocket with an exponentially growing density of states $D(E) 
\propto \exp(E/E_i)$, one finds that for $T < E_i$, the system is trapped in the 
local minimum at the bottom of the pocket, i.e. the high barriers of the pocket keep 
the system isolated from the rest of the energy landscape. But for $T > E_i$, the 
system  leaves the pocket with overwhelming probability, irrespective of the 
depth of the pocket. For a further discussion of exponential trapping and 
the competition among several exponential traps, see ref.\cite{Schon97b}.
 
If the assumption of thermalization within the pocket  holds true for $T 
\leq E_i$,   all quasi-equilibrium properties of the system can be 
calculated by the 
usual formulae of statistical mechanics, but with the sums over states restricted 
to a pocket. As the local density of states is available, we can calculate average 
energies and heat capacities.  The model specific heat was calculated for the 
example pockets shown in figure 3, and plotted as a function of $T$ in fig. 7.
As one would expect, $C_V$  shows a clear maximum at a temperature close to 
the average inverse growth factor of the local density of states. The height of this 
peak 
is the larger,  the larger the pocket is, and it is more pronounced the less the DOS 
deviates from an ideal exponential growth law. Note that deviations from a perfect 
exponential growth show up as additional features in $C_V$. In particular, a rapid 
exponential increase (with $E_i (1)$) of the DOS at low energies followed by a 
slower exponential growth (with $E_i (2) > E_i (1)$) (c.f. curves (b1) and (b2) 
in fig. 3b)) is reflected in a prepeak
at $T \approx  E_i (1)$ followed by the major peak at a temperature somewhat 
below $E_i(2)$ (curves (c) and (e) in fig. 7, respectively).		.
\footnote{This result agrees with the analytical calculation \cite{Schon97b} for the 
specific heat of a system restricted to a pocket, with an exponential $DOS$ of 
depth $D_i$ between the 
minimum and the top of the (exponential part of the) pocket and an inverse growth 
factor $E_i: C_V = 1$ for $T<<E_i$, $C_V = 1/T^2$ for $T>>E_i$, and $C_V = 
D_i/12E_i^2$ 
for $T = E_i$. Note that the behaviour of $C_V$ for $T >> E_i$ is a consequence of 
considering only the states with energies below the maximal lid $L_{max}$. If e.g. 
the exponential growth is followed at higher energies $E > L_{max}$ by a power law 
growth, $C_V$ approaches a constant value with increasing $T$ after peaking at 
$T \approx E_i$.} 
 
A qualitatively similar behaviour of the excess specific heat $C_p$ is observed in a 
large number of glass-forming systems, ranging from molecular and polymer 
systems to metallic and covalent glasses \cite{Gutzow95}. Thus, the exponential 
growth 
of the local DOS of the network could be responsible for the configurational entropy 
observed in experiment. As a consequence, the glass transition would be the result 
of the system experiencing exponential trapping \cite{Schon97b}, with the glass 
transition 
temperature $T_G \approx E_i$. In this context, one should point out that the 
high-
temperature tail ($T>E_i$) of the calculated specific heat of a pocket has no direct 
physical relevance when comparing the model with real glasses, since for $T > E_i$
the system would rapidly leave the pocket, and its equilibrium properties would no 
longer be dominated by the local density of states belonging to a single pocket.

This hypothesis of the thermodynamics of the glass transition being controlled by 
the trapping temperatures of locally ergodic, exponentially growing regions of the 
energy landscape of the glass raises the important question of the behaviour of the 
model 
system in the thermodynamic limit, $N_A,V \rightarrow \infty $ with $\rho = 
N_A/V = const.$ 
Clearly, the number of possible neighbours of a configuration grows to infinity, and 
similarly the growth factor of the local $DOS$, i.e., the trapping temperature 
$E_i$ goes 
to zero. This follows from the fact that due to the short range of the covalent 
interactions the energetic 
barriers in the system would be expected to grow only with $V^{(d-1)/d}$, 
 while the energy, as an extensive quantity,   grows proportionally  to $V$  itself. But one 
must not overlook the fact that there will be 
large entropic barriers preventing  the system from exploring this infinite set of 
neighbouring states even though these  are not  separated by unsurmountable energetic 
barriers. From a certain point on, the system size has grown to such enormous 
proportions that the dynamics is controlled by entropic barriers, i.e. one is no longer 
allowed to assume that on the relatively short time scales available for 
observation the system can e.g. "focus" all the thermal energy present in 
the network into a precise sequence of moves needed e.g. to cross some barrier to a 
neighbouring basin. Else, we would be on time scales where the glassy state can be 
transformed into the crystalline one, with the consequence that the configurational 
entropy vanishes, of course.

The existence of entropic barriers has important consequences for the dynamical 
behaviour of the system. Since the trapping temperature is a local equilibrium 
quantity of an exponentially growing region $\cal R$ of the energy landscape, it is 
necessary, in principle, to establish local ergodicity \cite{Schon97,Schon98b} 
within $\cal R$, at 
temperatures below the trapping temperature. Usually, energetic barriers serve to 
delimit such regions, but in the thermodynamic limit entropic barriers fulfill this 
task. However, visualisation and characterization of such entropic barriers is 
usually not straightforward.

The simplest picture of entropic barriers with regard to the energy landscape would 
be to assume that the configuration space of the excitations of the (infinite) 
system can be approximately separated into a direct product of independent 
subspaces. Each such subspace can be treated in analogy to the independent 
modes of e.g. vibrations, and would usually be visualized as a "cluster of atoms" 
within the network that has its individual excitation spectrum and corresponding 
density of states. Such clusters would be only weakly correlated with each other.
Since the actual number of states within a pocket of the landscape of such a 
cluster is small compared to the number of clusters in an infinite system, the 
distribution of energy throughout the system equals a Poisson distribution over the 
clusters. In particular, the entropy $S(E,V,N)$ 
becomes an extensive function of the number 
of clusters $N_C$: $S(E,V,N) = N_C S_C$, where $E_C = E/N_C,\;  V_C=V/N_C, \;  
$ and $N_{AC}=N_A/N_C)$. The quantity 
$S_C$ is the entropy of a single cluster, and for an exponential density of states
 is proportional to $E_C/E_i$, 
similarly to  the examples presented in this paper.
If one now assumes that the glass transition is a consequence of the energy 
landscape consisting of (possibly nested) locally ergodic pockets (e.g. the "clusters" 
discussed above) with exponentially growing densities of states, one would identify 
$T_G$ with the average inverse growth factor $E_i(V_C)$ of such regions. Using 
the simple growth law ($E_i \propto 1/V$) derived in the appendix, such an 
identification yields an 
estimate of the size of these clusters. Since in the independent cluster 
approximation clusters of size $V_C$ suffice to describe many of the 
thermodynamic 
properties of the networks, this establishes a reasonable network size for 
structural investigations at the level of network topology.

In the case of 2d-networks, the trapping temperature for a $20 \times 20$ system 
with $50 - 60$ atoms lies in the range of about $1 eV (\approx10^4 K)$. Since the 
glass temperature 
for covalent networks lies in the range of $0.1 eV (\approx10^3 K)$, the energy 
landscapes of networks with $N_A > 500$ might be suitable for a realistic description of 
properties of glasses.
While this number has been derived from the analysis of 2d-networks, preliminary results 
for 3d-networks indicate that the trapping temperature for e.g. networks with $40$ atoms lie at about $1.2$ eV/atom, leading again to an estimate of the needed network size of about $N_A > 500$.

\section{Conclusions}

In this paper, we have presented a first in-depth study of the microscopic energy 
landscape of glasses at the level of their network topology. Such amorphous 
networks show an exponential growth of many important quantities, in particular 
the local density of states and minima, $D(E;L)$ and $D_M(E;L)$ respectively, 
the accessible state space volume $N(L)$ and the accessible neighbor-minima 
$M(L)$. This 
growth leads to exponential trapping that can explain the occurrence of a glass 
transition in such systems. In particular, the peak in the excess specific heat at 
the glass transition, that is often observed in experiments, follows directly from the 
exponential growth law. 

This type of local exponential growth appears to be a common feature of many 
complex systems. Similar behaviour has been found for polymers
 \cite{Schon97,Schon98}, spin 
glasses \cite{Sibani94,Sibani98b,Hoffmann98}, combinatorial optimisation problems \cite{Sibani93}, 
crystalline solids 
\cite{Putz98,Wevers98} and, in preliminary results by the present authors,
 for three-dimensional random networks.

It is expected that these results can form a more solid basis for phenomenological 
models of complex systems. An important open question is the issue of entropic 
barriers, since they control the effective size of locally ergodic pockets in large 
systems. Knowing this size would allow the direct quantitive comparison of the 
calculations with experimental data; e.g. one could predict the glass transition 
temperature and its dependence on cooling rates.
 
\noindent{\bf Acknowledgments}\\
We gratefully acknowledge funding by the DFG via the SFB 408.
This work was partly supported by a block grant from  
the Danish Statens Naturvidenskabelige Forskningsr\aa d.

\section{Appendix}  
For the qualitative and semi-quantitative analysis of the network glasses, the 
following very simple free-volume-model that is based on some rough assumptions 
about the nature of the energy landscape can be helpful: 
1. Starting from each configuration,  $2dN_A$ new configurations can be 
generated by moving each atom to one of the  $2d$ neighboring sites on the lattice 
($d$ is the dimension of the lattice).
2. The interactions are highly local, such that the typical increase in energy 
$E_M$ of an acceptable move, i.e. one which leads to a configuration below the 
energy lid, is essentially independent of the number of atoms in the simulation 
cell.
 The typical increase in the energy/atom for a system with $N_A$ atoms when
  an acceptable move is performed is thus $E_M/N_A$.
3. As long as the states belong to the fast-growing region of the "pocket",
 the number of downhill moves among the acceptable moves will be very small. 
 Thus we assume that a given 
move will result in a new configuration with an energy increase $\delta E \simeq E_M$ 
with a probability $f(\rho,N_A)$. In particular,
 the probability $f_-$ for a downhill move should be very small compared to $f$,
  $f_- << f$.
Thus, the number of configurations
$N(E+\delta E)$ with an energy below $E+\delta E$, 
 is proportional to the number of configurations below the energy 
$E$, $N(E)$:
\begin{equation}
 N(E+\delta E) \propto N(E)(2 d N_A f)
\label{A1}
\end{equation}
 from which it easily follows that 
\begin{equation} 
 N(E )  = N_0 \exp(\alpha E), 
 \label{A3}
 \end{equation}  					 
with 
$\alpha = 1/E_i = (1/E_M)(2 d N_A f -1) \approx 2 d N_A f/E_M$.
The last approximation should be acceptable in the limit $N_A \rightarrow 
\infty$ , as long as $f$ varies only weakly with $N_A$ (for fixed $\rho$). 

It seems  reasonable to assume that $f$ depends on the density of the system, and that  
it  increases with the free volume per atom $v_f = V_f/N_A$ in the system, where the 
free volume $V_f$ is given  in terms of the volume per atom $V_A$  as 
$V_f = V - N_A V_A$.  Thus  we get, for some constant $c$ 
\begin{equation}
f = c  v_f = c (V - N_A V_A)/N_A = c (V/N_A)(1 - \rho V_A),
\label{A6}
\end{equation} 		 
and therefore 
\begin{equation}
\alpha = 1/E_i \approx 2 d N_A f/E_M \propto V(1 - \rho V_A) 
\label{A7}		
\end{equation}		 
Equation \ref{A6}  shows  that $f$ increases with decreasing density for constant 
volume. Thus  $E_i$ also decreases with density, and secondly, for constant 
density 
\begin{equation}
E_i \propto 1/V .
\label{A8}
\end{equation}					 
This final result agrees quite  well with the observed behaviour.

Note that according to eq.\ref{A7},  $E_i$ should actually decrease as $1/V(1 - \rho V_A)$.
 In order to check this behaviour, we need to determine $V_A$. This value 
corresponds to the number of lattice points that the atom would occupy in an 
energetically favourable configuration (recall that we are investigating the pockets 
around deep-lying minima). For an approximately hexagonal arrangement on the 
square lattice, we find $V_A = 6$. If one plots $Y ( = N_A E_i)$ against $X ( = 1/(1 - 
\rho V_A))$ for all available volumes $V = S^2 (S = 10,...,20)$ one finds that the 
data roughly follow straight lines as suggested by eq.\ref{A7} \cite{Schon97}.

Note that there exists only one free parameter $(c/E_M)$ in this simple free 
volume model. The number of accessible 
states $N(L)$ estimated by  this model exhibits an average exponential growth 
similar to 
$D(E)$ (c.f. fig. 2), and the qualitative dependence of the growth factors on 
$V=S^2$ and 
$\rho$ should be essentially the same for $D(E)$ and $N(L)$, unless e.g. there is a 
high degree of degeneracy in the ground state of the pocket. But even for polymer 
systems, where such a degenerate ground state occurs more frequently, an 
analogous free volume model roughly describes the dependence of the growth 
factors on system size \cite{Schon97,Schon98}.

\begin{figure}[t]
\centerline{\psfig{figure=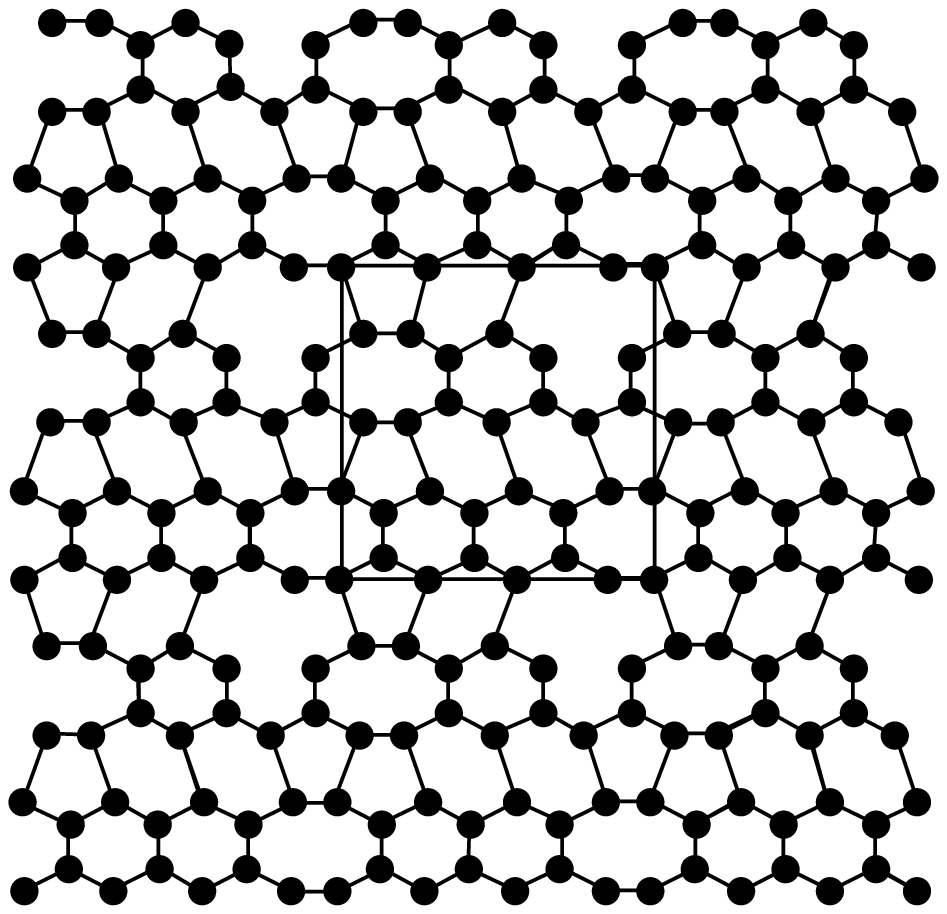}}
\vspace{1cm}
\caption{
A typical network configuration belonging to a low-energy metastable state
 in a system of $27$ atoms with a unit  cell of linear size $14$. 
} 
\end{figure}
\begin{figure}[t]
\vspace{-2cm}
\centerline{\psfig{figure=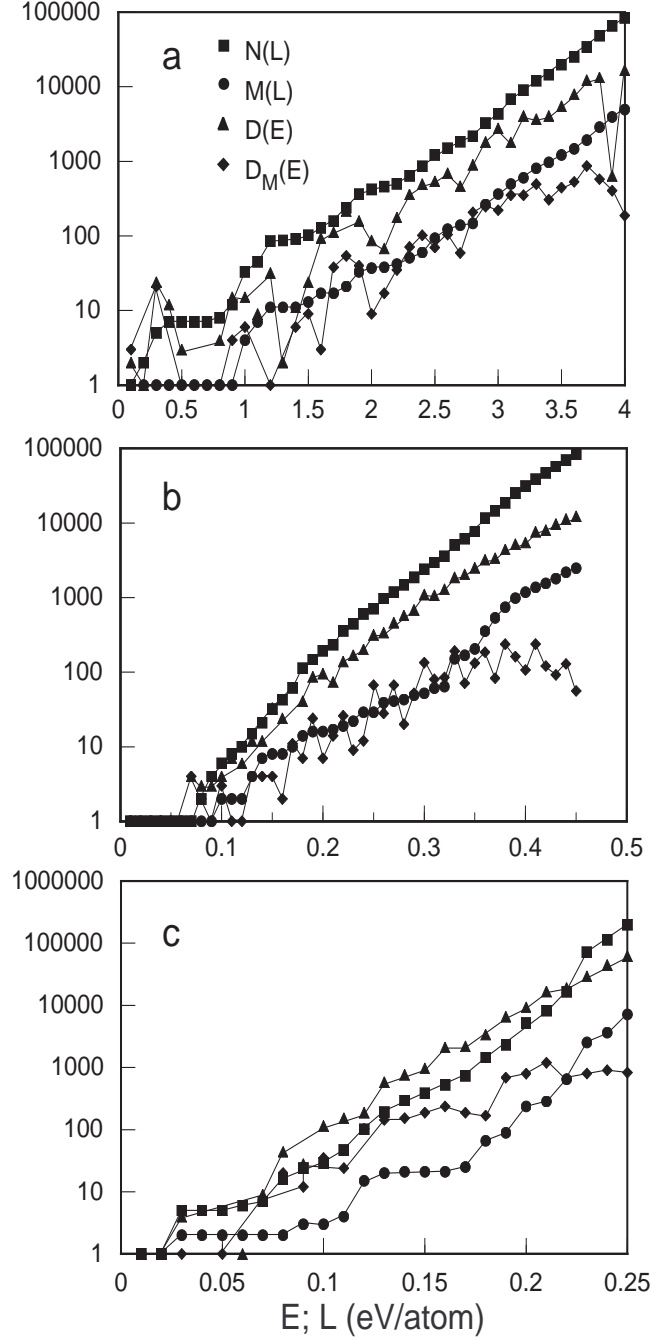,height=18cm,width=9.5cm}}
\vspace{1cm}
\caption{
Number of accessible states $N(L)$ (squares), accessible minima 
$M(L)$ (circles), density of states $D(E)$ (triangles) and density of 
minima $D_M(E)$ ( diamonds) within a pocket, as a function of $L$ 
[eV/atom] and $E$ [eV/atom], respectively. Data are shown 
 for three different pockets: a) $(N_A = 22; S = 12)$, 
b) $(N_A=36;S=16)$ and c) $(N_A = 53; S=20)$. 
} 
\end{figure}
\newpage
\begin{figure}[t]
\vspace{-4cm}
\centerline{\psfig{figure=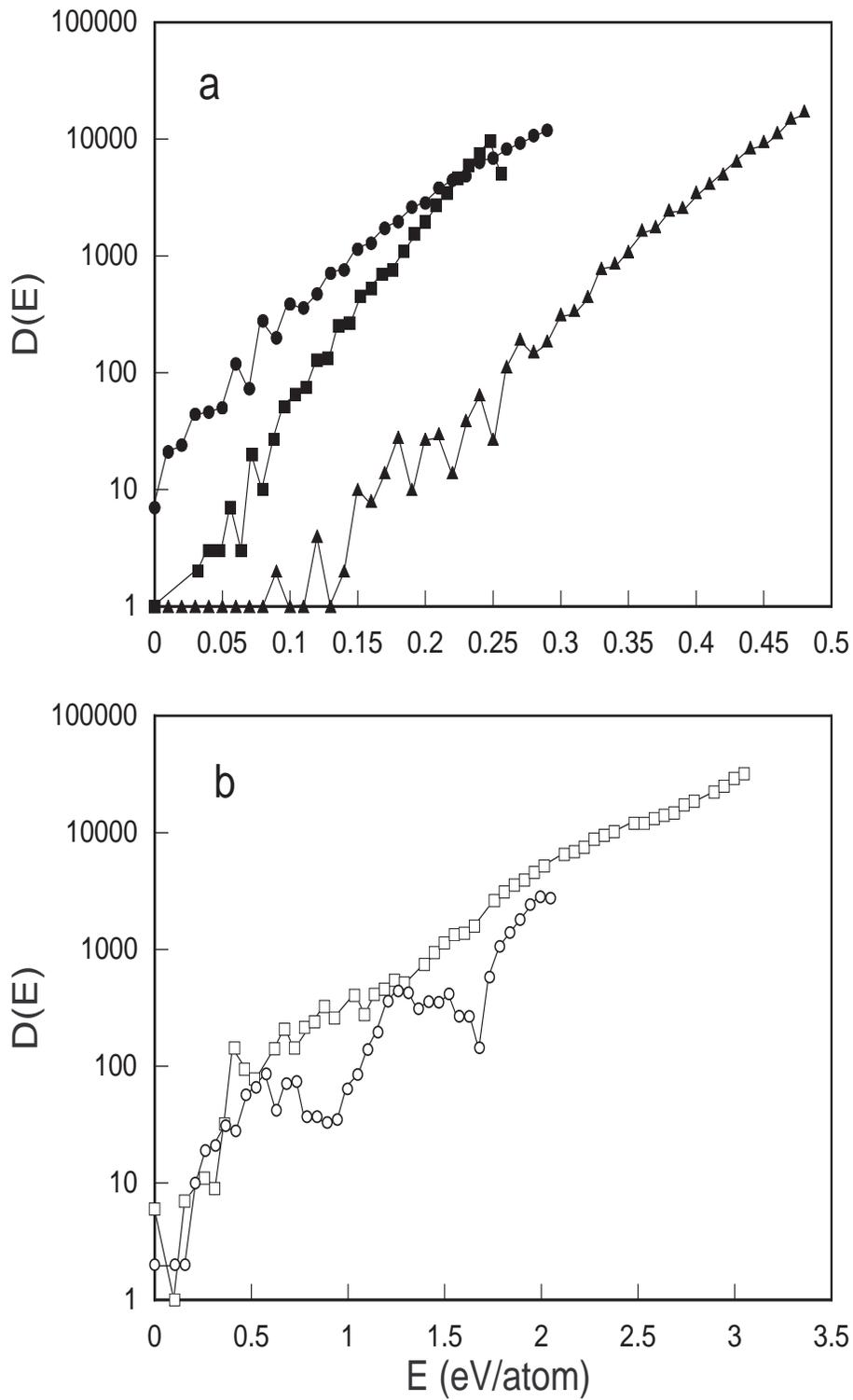,height=18cm,width=13cm}}
\vspace{1cm}
\caption{
Local  density of states $D(E)$ vs. $E$ [eV/atom]  for several 
representative systems: a):  black squares (a1): $(N_A=46;S=18)$, black circles (a2):
 $(N_A=41;S=17)$,   
and black triangles (a3): $(N_A=34;S=16)$. b): white squares (b1): $(N_A=13;S=10)$ 
and white circles (b2): 
$(N_A=21;S=12)$. Note the curves belonging to the larger systems shown in a) are much
steeper than those belonging to the smaller system 
 shown in b). 
} 
\end{figure}
 \begin{figure}[t]
\centerline{\psfig{figure=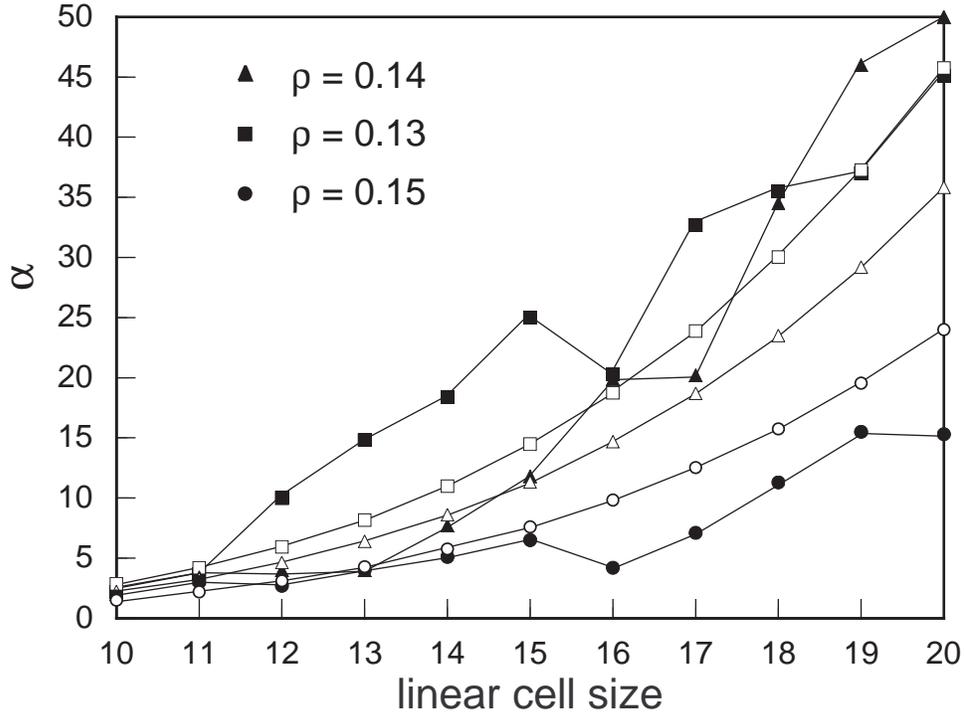}}
\vspace{0.5cm}
\caption{
 The quantity $\alpha = 1/<E_i>$ [1/(eV/atom)] for different densities:
black squares: $\rho = 0.13$, black triangles: $\rho = 0.14$   and 
black circles: $\rho = 0.15$.
The white symbols show the corresponding theoretical predictions
obtained through eq.\ref{A7}, with $V_A = 6$. 
The average over the inverse growth factors
$E_i$ is performed  over all pockets in systems of the same density  
and cell size.   
} 
\end{figure}    
\begin{figure}[t]
\centerline{\psfig{figure=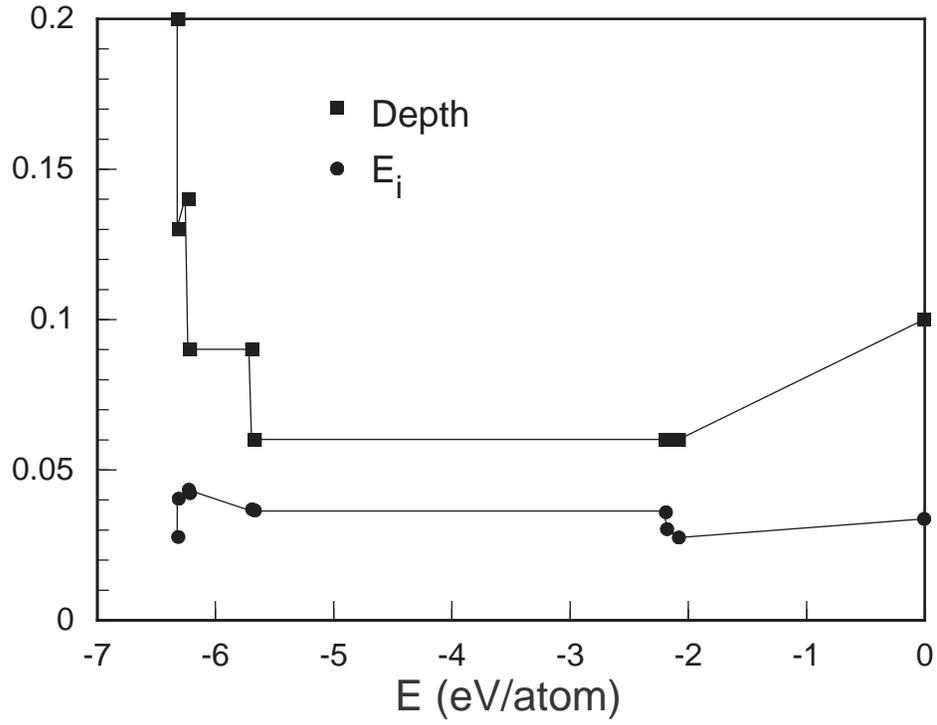}}
\vspace{0.5cm}
\caption{ Depth  (squares) and inverse growth factors $E_i$, both 
quantities expressed in  
$[eV/atom]$  for several sample pockets as a function of the
 lowest energy of the pocket. The system considered has
 $(N_A = 39; S = 17)$. 
} 
\end{figure}   
\newpage
\begin{figure}[t]
\centerline{\psfig{figure=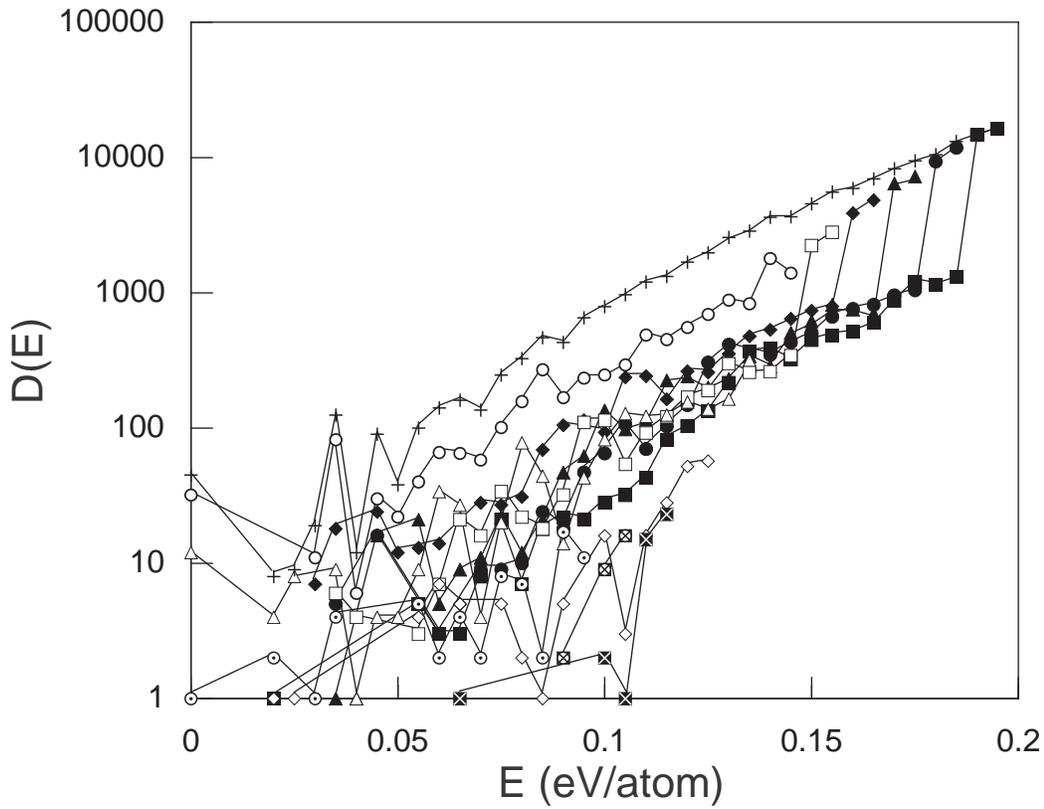}}
\vspace{0.5cm}
\caption{
Local densities of states of several sub-basins of the `ground state' pocket in the 
system $(N_A = 39; S = 17)$, which appear successively as the lid is raised. 
 The top curve is the local density of states of the whole pocket. Note that the
 exponential trend is common to  all sub-basins.
} 
\end{figure}
\newpage 
\begin{figure}[t]
\centerline{\psfig{figure=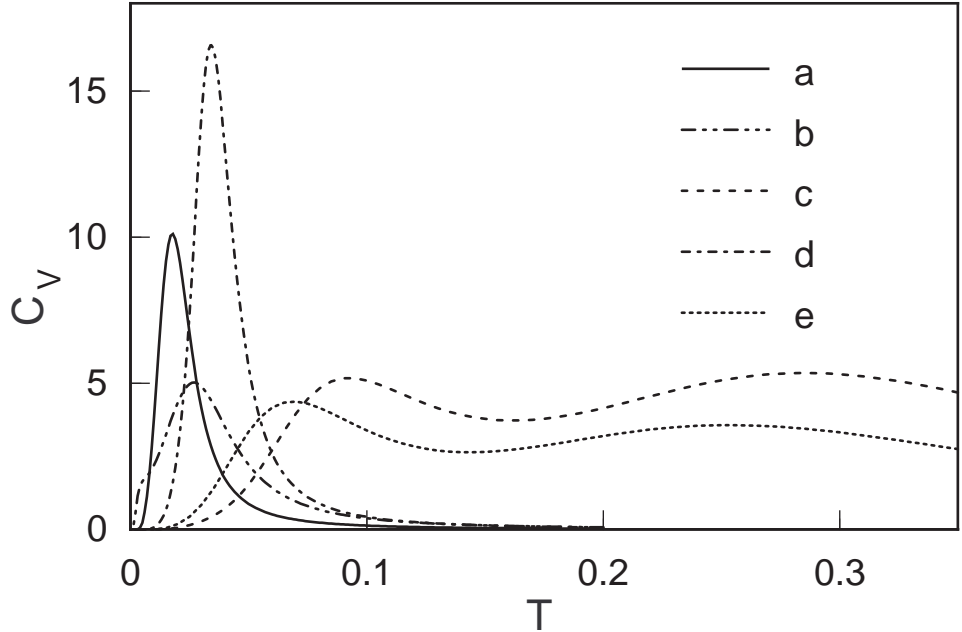}}
\vspace{0.3cm}
\caption{
Specific heat in units of $[k_B]$ vs. temperature [eV/atom] calculated
for several representative  examples: 
a) $(N_A=46;S=18)$, b) $(N_A=41;S=17)$, c) $(N_A=13;S=10)$, d) 
$(N_A=34;S=16)$, and e) $(N_A=21;S=12)$.  The three peaks in 
plots  a), b) and d) are located at temperatures which match the 
inverse growth factors $E_i$ of the exponential density of states
shown in Fig. 3a), (a1), (a2) and (a3), respectively. 
The double peaks of plots (c) and (e) correspond to the
two different slopes characterizing the densities of states shown in 
Fig. 3b), (b1) and (b2), respectively.  
} 
\end{figure}

\end{document}